\documentclass[12pt]{article}
\usepackage{graphicx}

\def\Red{}
\def\Black{}
\def\Blue{}

\usepackage{epsf}

\newcommand{\GeV}{~{\rm GeV}}

\newcommand{\be}{\begin{equation}}
\newcommand{\ee}{\end{equation}}
\newcommand{\bea}{\begin{eqnarray}}
\newcommand{\eea}{\end{eqnarray}}
\newcommand{\beq}{\begin{equation}}
\newcommand{\eeq}{\end{equation}}
\newcommand{\ba}{\begin{array}}
\newcommand{\ea}{\end{array}}
\newcommand{\beqa}{\begin{eqnarray}}
\newcommand{\eeqa}{\end{eqnarray}}

\newcommand{\cO}{{\cal O}}

\newcommand{\no}{\nonumber}
\newcommand{\lsim}{\stackrel{<}{_\sim}}

\newcommand{\Mp}{\Lambda_{\rm Pl}}
\newcommand{\md}[1]{\langle#1\rangle}
\newcommand{\Det}{\mathop{\rm Det}}

\newcommand{\SDet}{\mathop{\rm SDet}}
\newcommand{\phib}{h}
\newcommand{\Dsl}{D\hspace{-1.5ex}/}
\newcommand{\ds}{\partial\hspace{-1.2ex}/}

\def\npb#1#2#3{    {\it Nucl. Phys. }{\bf B #1} (#2) #3}
\def\plb#1#2#3{    {\it Phys. Lett. }{\bf B #1} (#2) #3}

\def\prd#1#2#3{    {\it Phys. Rev. }{\bf D #1} (#2) #3}

\def\prep#1#2#3{   {\it Phys. Rep. }{\bf #1} (#2) #3}
\def\prl#1#2#3{    {\it Phys. Rev. Lett. }{\bf #1} (#2) #3}

\def\ppnp#1#2#3{   {\it Prog. Part. Nucl. Phys. }{\bf #1} (#2) #3}

\def\zpc#1#2#3{    {\it Z. Phys. }{\bf C #1} (#2) #3}

\def\ibid#1#2#3{   {\it ibid. }{\bf #1} (#2) #3}

\begin{document}

\thispagestyle{empty}
\begin{flushright}
CERN--TH/2001--092 \\
LNF-01/014(P)\\
GeF/TH/6-01\\
IFUP--TH/2001--11 \\
hep-ph/0104016
\end{flushright}
\vskip 3.0 true cm 

\begin{center}
{\Large\bf\Red On the metastability of the Standard Model vacuum\Black}\\
[25 pt]
{\bf Gino Isidori\footnote{On leave from INFN, Laboratori Nazionali di
 Frascati, Via Enrico Fermi 40, I-00044 Frascati, Italy.},
 Giovanni Ridolfi\footnote{On leave from INFN, Sezione di Genova,
 Via Dodecaneso 33, I-16146 Genova, Italy.}
 and Alessandro Strumia}\footnote{On leave from
 Dipartimento di Fisica, Universit\`a di Pisa and INFN, Sezione di Pisa,
 Italy. }\\ [10pt]
{\em Theory Division, CERN, CH-1211 Geneva 23, Switzerland}

\vskip 2.0 true cm
\Blue{\bf Abstract} \\
\end{center}
\noindent
If the Higgs mass $m_H$ is as low as suggested by present experimental
information, the Standard Model ground state might not be absolutely
stable.  We present a detailed analysis of the lower bounds on $m_H$
imposed by the requirement that the electroweak vacuum be sufficiently
long-lived.  We perform a complete one-loop calculation of the
tunnelling probability at zero temperature, and we improve it by means
of two-loop renormalization-group equations. We find that, for
$m_H=115$~GeV, the Higgs potential develops an instability below
the Planck scale for $m_t>(166\pm 2) \GeV$, but the electroweak vacuum
is sufficiently long-lived for $m_t < (175\pm 2) \GeV$.

\def\thefootnote{\arabic{footnote}}
\setcounter{footnote}{0}
\setcounter{page}{0}

\setcounter{equation}{0}
\renewcommand{\theequation}{\thesection.\arabic{equation}}
\newcommand{\mysection}[1]{\section{ #1 }\setcounter{equation}{0}}

\newpage
\Black
\mysection{Introduction}

If the Higgs boson if sufficiently lighter than the top quark,
radiative corrections induced by top loops destabilize the electroweak
minimum and the Higgs potential of the Standard Model (SM) becomes
unbounded from below at large field values. The requirement that such
an unpleasant scenario be avoided, at least up to some scale $\Lambda$
characteristic of some kind of new physics~\cite{stab,sher}, leads to
a lower bound on the Higgs mass $m_H$ that depends on the value of
the top quark mass $m_t$, and on $\Lambda$ itself. The most recent
analyses of this bound~\cite{stab_new}, performed after the discovery
of the top quark, led to the conclusion that if the Higgs boson was clearly
observed at LEP2 (or if $m_H\lsim 100$~GeV), then new physics would
have to show up well below the Planck scale, $\Mp=1.2\times
10^{19}$~GeV, in order to restore the stability of the electroweak
minimum. We now know that $m_H$ must be larger than about
113~GeV~\cite{LEPbound}. If $m_H$ lays just above this bound, as
hinted by direct searches~\cite{Lep2_ev} and consistently with
electroweak fits~\cite{Higgs@Lep}, absolute stability up to the Planck
scale is possible, provided $m_t$ is close to the lower end of its
experimental range~\cite{HBNielsen}. For $m_t$ around its central
value, the SM vacuum may not be absolutely stable, but still
sufficiently long-lived with respect to the age of the universe. Motivated
by these observations, we decided to reanalyse in detail the lower
limits on $m_H$ imposed by the condition of (meta-)stability of the
electroweak minimum.

We assume that no modifications to the Standard Model occur at scales
$\Lambda$ smaller than the Planck scale. In general, field-theoretical
modifications invoked to stabilize the SM potential at scales
$\Lambda\ll \Mp$, such as supersymmetry, or the introduction of extra
scalar degrees of freedom, induce computable corrections of order
$\Lambda^2$ to the squared Higgs mass, thereby forcing $\Lambda$ to be
of the order of the electroweak scale by naturalness arguments. On the
other hand, it cannot be {\em a priori} excluded that the uncomputable
gravitational corrections of order $\Mp^2$ vanish.

Three different classes of bounds have been discussed in the
literature~\cite{sher}:
\begin{enumerate}
\item[i)] absolute stability; 
\item[ii)] stability under thermal fluctuations in the hot universe;
\item[iii)] stability under quantum fluctuations at zero temperature.
\end{enumerate}
The condition of absolute stability is the most stringent one.
However, although appealing from an aesthetic point of view, this
constraint is not demanded by any experimental observation: it is conceivable
that we live in an unstable vacuum, provided only that it is not
``too unstable''.  The condition (ii) --- less stringent than (i) and
more stringent than (iii) --- relies on the assumption that the early
universe passed through a phase of extremely high temperatures
(the most stringent bounds are obtained for $T \sim\Mp$).  Although
plausible, this is just an assumption; so far, it has been indirectly
tested only for temperatures up to few MeV. A na\"{\i}ve
extrapolation of big-bang cosmology by $\sim 20$ orders of magnitude
in temperature would not only give a bound on the SM Higgs mass; it
would also exclude various popular unified, supersymmetric or
extra-dimensional models, because of over-abundance of
monopoles, gravitinos, gravitons, respectively.

Finally, the requirement of sufficient stability under quantum
fluctuations at zero temperature gives the less stringent bounds, but
does not rest on any cosmological assumptions.  The only cosmological
input required is an approximate knowledge of the age of the universe
$T_U$; the bound is formulated by requiring that the probability of
quantum tunnelling out of the electroweak minimum be sufficiently
small when integrated over this time interval.  In this work we will
mainly concentrate on this scenario.

The probability that the electroweak vacuum has survived quantum 
fluctuations until today is given, in semi-classical approximation,
by~\cite{coleman_sc}
\beq
p \approx (T_U/R)^4  e^{-S_0}~, 
\label{eq:1}
\eeq
where $S_0$ is the Euclidean action of the {\em bounce}, the
solution of the classical field equations that interpolates between
the false vacuum and the opposite side of the barrier, and $R$ is a
dimensional factor associated with the characteristic size of the bounce.  
The main purpose of this paper is to reduce the theoretical
uncertainties in the above result, performing a complete one-loop
calculation of the action functional around the bounce configuration
\cite{coleman_1L}. As we shall show, this calculation allows us to
unambiguously fix the pre-exponential factor and the finite
corrections at the one-loop level, and also to consistently resum (by
means of renormalization group equations) the sizeable logarithmic
corrections appearing in the exponential factor.

The paper is organized as follows: in Section~2 we shall briefly
recall the semi-classical result for the tunnelling rate, applied to
the case of the SM Higgs. In Section~3 we shall discuss the general
properties of the one-loop formula, emphasizing the differences with
the semi-classical one. Section~4 contains all the technical details
of the calculation, whereas the numerical bounds on the Higgs mass are
presented in Section~5. Finally we summarize our results in Section~6.

\mysection{Tree-level computation of the tunnelling rate}

The Standard Model contains a complex scalar doublet $\phi$ with
hypercharge $-1$,
\beq
\phi = \left[ \ba{c} (H+iG)/\sqrt{2} \\ G^- \ea \right]~,
\eeq
and tree-level potential
\beq
\label{eq:SMV}
V(\phi)=m^2 |\phi|^2 + \lambda |\phi|^4
=\frac{1}{2} m^2 H^2 + \frac{1}{4}\lambda H^4+\ldots
\eeq
where the dots stand for terms that vanish when $G,G^-$ are set to
zero. The neutral component $H$ is assumed to acquire a non-vanishing
expectation value $\md{H}=v$. With this normalization, $v=
(G_F\sqrt{2})^{-1/2}= 246.2 \GeV$, and the mass of the single physical
degree of freedom $H$ is $m^2_H=V''(H)|_{H=v}=2 \lambda v^2$. As is
well known, for $H \gg v$ the quantum corrections to $V(H)$ can be
reabsorbed in the running coupling $\lambda(\mu)$, renormalized at a
scale $\mu \sim H$. To good accuracy, $V(H \gg v )=
\frac{1}{4}\lambda(H) H^4~$ and the instability occurs if, for some
value of $H$, $\lambda(H)$ becomes negative. Since, for $m_H$ larger
than 100~GeV, this occurs at scales larger than $10^5$~GeV, we shall
neglect the quadratic term $m^2H^2/2$ throughout the paper.

In general, the bounce~\cite{coleman_sc} is a solution $H=\phib(r)$ of
the Euclidean equations of motion that depends only on the radial coordinate
$r^2\equiv x_\mu x_\mu$:
\beq
-\partial_\mu \partial_\mu \phib+V'(\phib ) 
=-\frac{d^2 \phib }{dr^2}-\frac{3}{r}\frac{d\phib }{dr}+V'(\phib)=0~,
\eeq
and satisfies the boundary conditions
\beq
\phib'(0)=0~,\qquad \phib(\infty) = v\to 0~.
\eeq
We can perform a tree-level computation
of the tunnelling rate with $\lambda<0$. This leads to 
\begin{equation}
\label{eq:fubini}
\phib(r) = \sqrt{\frac{2}{|\lambda| }}\frac{2R}{r^2+R^2}~,\qquad S_0[\phib] 
=\frac{8\pi^2}{3|\lambda|}~,
\end{equation}
where $R$ is an arbitrary scale. At first sight, the approximation of
taking $V(h)=\lambda h^4/4$ may appear rather odd, since the unstable
vacuum configuration $h=0$ corresponds to the maximum of the
potential. However, this is not a problem within quantum field
theory, since the tunnelling configuration requires a non-zero kinetic
energy (the bounce is not a constant field configuration) and is
therefore suppressed even in the absence of a potential
barrier~\cite{LeeWeinberg}. The
SM potential is eventually stabilized by unknown new physics around
$\Mp$: because of this uncertainty, we cannot really predict what
will happen after tunnelling has taken place. Nevertheless, a computation
of the tunnelling rate can still be performed~\cite{coleman_sc}.

The arbitrary parameter $R$ appears in the expression of the bounce
since, because of our approximations, the potential is
scale-invariant: at this level, there is an infinite set of bounces of
different sizes that lead to the same action.

\begin{figure}[t]
\begin{center}
\includegraphics{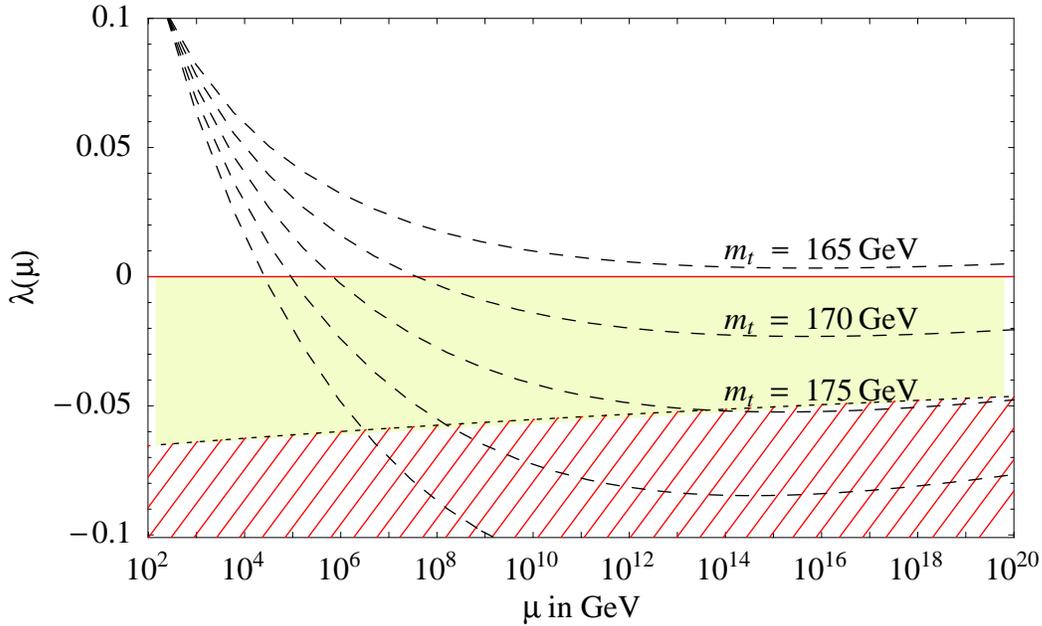}
\caption{\em Running of the quartic Higgs coupling for 
  $m_H = 115\GeV$ and $m_t= 165, 170, 175,180$ and $185\GeV$
  $[\alpha_s(m_Z)=0.118]$. Absolute stability $[\lambda > 0]$ is
  still possible if $m_t < 166\GeV$. The hatched region indicates the
  metastability bound.}
\label{fig:lambda}
\end{center}
\end{figure}

Substituting the bounce action (\ref{eq:fubini}) in Eq.~(\ref{eq:1}),
the condition $p < 1$ for a universe about $10^{10}$ years old is
equivalent to $\lambda > -0.065/(1 - 0.01 \ln R v)$, i.e.\ 
$\lambda$ cannot take too large negative values. The bound on
$\lambda$ can be translated into a lower bound on $m_H$ taking into
account the renormalization-group evolution (RGE) of $\lambda(\mu)$
(see Fig.~\ref{fig:lambda}).
At this stage, however, there is clearly a large theoretical ambiguity
due to the scale dependence. Which values of $R$ and of the RGE scale
$\mu$ should one use? As we shall see in the next section, both
ambiguities are solved by performing a complete one-loop calculation of
the tunnelling rate.

\mysection{Quantum corrections to the tunnelling rate}

The procedure to compute one-loop corrections to tunnelling rates
in quantum field theory has been described by Callan and Coleman
\cite{coleman_1L}, following the work of Langer~\cite{Langer} in
statistical physics. At this level of accuracy, the tunnelling
probability per unit four-dimensional volume $V$ can be written
as
\beq
\label{eq:S1}
{p \over V } = { e^{-S_1[\phib]} \over V }
 = \frac{S^2_0[\phib]}{4\pi^2} \,\left|
\frac{\SDet' S_0''[\phib]}{\SDet S_0''[0]} \right|^{-1/2}e^{-S_0[\phib]}~,
\eeq
where $\phib$ still denotes the tree-level bounce and $S_0$ ($S_1$) is
the tree-level (one-loop) action functional. Here $\phib=0$ indicates
the false (electroweak) vacuum; we assume that the potential has
been shifted so that $S_1[0]=0$; $S''_0$ denotes double functional
differentiation of $S_0$ with respect to the various fields; $\Det$ is
the functional determinant, and $\SDet \equiv \Det$ or $\SDet \equiv
1/\Det^2$, depending on whether it acts on boson or fermion fields.

When evaluated with a constant field configuration, $S_1$ is simply
given by the usual one-loop effective potential. Computing
$S_1[\phib]$ is a much harder task because the
bounce~(\ref{eq:fubini}) is not a constant field configuration.
Furthermore, unlike the constant-field case, there are quantum
fluctuations that correspond to translations of the bubble. The `prime'
on $\SDet' S_0''[\phib]$ in Eq.~(\ref{eq:S1}) indicates that these
fluctuations, corresponding to zero modes, have been explicitly
removed from the functional determinant. In this way the result
acquires a dimensional factor that will be compensated by the
integration over the volume of the universe.

Fortunately, in order to compute the one-loop corrections to the
tunnelling rate we do not need to find the field configuration
$\phib_1$ that extremizes the full one-loop action: we only need to
compute $S_1[\phib]$, where $\phib$ is the field configuration that
extremizes $S_0$ and has the simple form in Eq.~(\ref{eq:fubini}).
The difference between $S_1[\phib]$ and $S_1[\phib_1]$ is a two-loop
correction.

\begin{figure}[t]
\begin{center}
\includegraphics{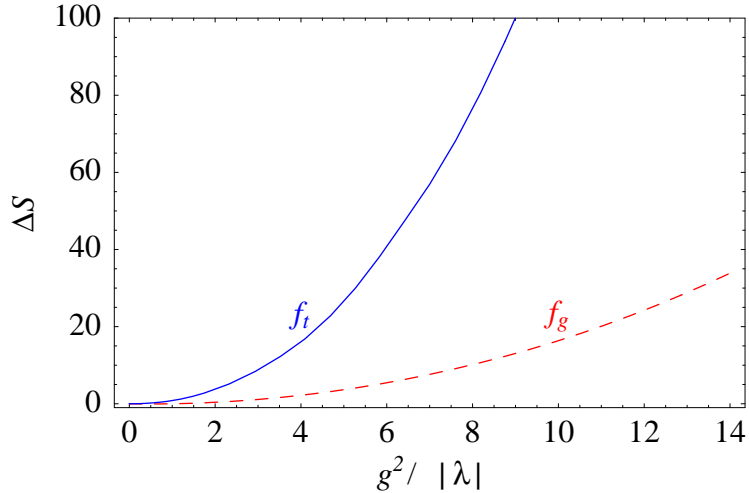}
\caption{
      \em Numerical results for the subtracted part of the correction
      to the action from top loops $[f_t(g_t^2/|\lambda|)$, solid
      curve$]$ and gauge boson loops $[f_g(g^2/|\lambda|)$, dashed
      curve$]$. Relevant values of $g^2/|\lambda|$ are $4\div 8$.}
\label{fig:ft}
\end{center}
\end{figure}

In our case, the main effect of quantum fluctuations is the breaking of
scale invariance of the tree-level potential. As we shall show
explicitly, this implies that bounces with different $R$, which have
the same action at the semi-classical level, turn out to have a
one-loop action roughly given by $S_1[\phib]\sim
8\pi^2/(3|\lambda(1/R)|)$. At the same time, for these configurations
the dimensional factor due to the zero eigenvalues turns out to be of
$\cO(R^{-4})$. In this way the two scale ambiguities of the
semi-classical result are completely resolved. Indeed, the complete
result for the tunnelling probability at one loop can be written as
\beq
\label{eq:p1L}
p = \max_R \frac{V_U }{R^4} 
\exp\left[ -\frac{8\pi^2}{ 3 \left| \lambda(\mu)\right|} 
-\Delta S \right] ~,
\eeq
where
\beqa
\label{eq:DeltaS}
\Delta S &=&-\frac{2}{3}(5+6L)+\frac{g_t^2}{6|\lambda|}(13+12L)
+\frac{g_t^4}{6\lambda^2}(5+6L)
-\frac{2g_2^2 + g_Z^2}{12|\lambda|}(7+6L)+\no\\
&&
-\frac{2g_2^4 + g_Z^4}{32\lambda^2}(1+2L)
+ f_h(\lambda) - 
f_t\Big(\frac{g_t^2}{|\lambda|}\Big)+
f_g\Big(\frac{g_Z^2}{|\lambda|}\Big)+2 
f_g\Big(\frac{g_2^2}{|\lambda|}\Big)~.\label{eq:risultato}
\eeqa
Here $V_U \sim T_U^4$, $L=\ln(R\mu e^{\gamma_{\rm E}}/2)$, $g_t$ is
the top Yukawa coupling, and $g_2^2$, $g_Z^2\equiv g_2^2+g_Y^2$ are
the weak gauge coupling constants, defined at tree level by $m_t=g_t
v/\sqrt{2}$, $m_W^2=g_2^2 v^2/4$, $m_Z^2=g_Z^2 v^2/4$. All couplings
are renormalized at the RGE scale $\mu$. The numerical values of the
functions $f_g$, $f_t$ are plotted in Fig.~\ref{fig:ft}, whereas
$f_{h}$ is given in Eq.~(\ref{eq:f_h}). The $L$ terms cancel the
$\mu$ dependence of $\lambda$ in the leading semi-classical term. If
one chooses a value of $\mu \sim 1/R$, such that $L\sim 0$, the
typical correction to the action is of $\cO(g_t^4/\lambda^2) \sim 10$, to be
compared with the leading term of order 100.

In previous analyses (see e.g.\ Ref.~\cite{QE}) a full one-loop
computation of the tunnelling rate was never performed, and the
semi-classical result was improved by considering only
quantum corrections to the effective potential, or to the running of
$\lambda$. This procedure leads to a correct estimate of the leading
logarithmic corrections to the action, but the finite terms of the
calculation are not under control. Within this approximation
the use of two-loop RGE equations does not improve the accuracy of
the calculation.
On the other hand, a consistent implementation of two-loop RGE equations for
$\lambda(\mu)$ is possible starting from Eq.~(\ref{eq:p1L}).

\begin{figure}[t]
\begin{center}
\includegraphics{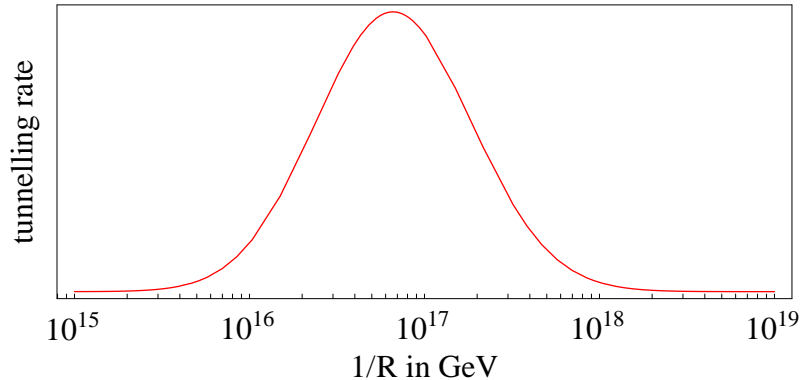}
\caption{\em Contribution to the tunnelling rate (in arbitrary units)
  from bubbles of different $R$, as a function of $1/R$,
  for $m_H = 115\GeV$ and $m_t = 175\GeV$.}
\label{fig:fR}
\end{center}
\end{figure}

In Fig.~\ref{fig:lambda} we plot the evolution of $\lambda(\mu)$ as
obtained by integrating the two-loop RGE equations of $\lambda$, the
top Yukawa coupling $g_t$ and the three gauge couplings $g_i$
\cite{Ford} for $m_H=115\GeV$ and some reference values of the pole
top mass $m_t$.\footnote{The initial values of $\lambda$ and and $g_t$
  have been related to the values of $m_H$ and $m_t$ using the
  matching conditions given in~\cite{SZ} and~\cite{HK}, respectively.
  The discussion about the uncertainties involved in this estimate of
  $\lambda(\mu)$ is posponed to Section~5.} For comparison we also
show the lower bound on $\lambda$ derived from Eq.~(\ref{eq:p1L}),
imposing the condition $p<1$ and assuming $V_U =(10^{10}~{\rm yr})^4$.
As can be noticed, the evolution of $\lambda$ crosses the
metastability bound (i.e.~the tunnelling rate becomes too high) for
values of $\mu$ much larger than the electroweak scale. This implies
that our approximation of neglecting the $\cO(v^2)$ quadratic term in
the tree-level potential is very good, since the critical bounces are
those with a size much smaller than $1/v$. It is also important to
notice how the lower bound on $\lambda$ increases as a function of the
RGE scale (or~of $1/R$). This effect is due to the pre-exponential
factor in Eq.~(\ref{eq:p1L}), scaling like $R^{-4}$, which we have been
able to determine from the one-loop computation. It is
important to notice that, for the experimentally interesting values of
$m_H$ and $m_t$, the tunnelling rate is dominated by bubbles with
$1/R$ about two orders of magnitude below $\Mp$, as can be seen
in Fig.~\ref{fig:lambda} or, more clearly, in Fig.~\ref{fig:fR}.
Therefore the metastability bound on $m_H$
does not depend on the unknown physics around $\Mp$.

\mysection{Explicit computation of the one-loop action}

\subsection{General strategy}
\label{subs:GS}

The central point in the computation of the tunnelling
probability, Eq.~(\ref{eq:S1}), is the evaluation of
ratios of functional determinants:
\beq
\label{eq:detratio}
\frac{\Det S_0''[\phib]}{\Det S_0''[0]} 
\eeq
in the various sectors of the theory. This requires solving
eigenvalue equations of the type 
\beq
\label{eq:eigeq}
S_0''[\phib]\,\psi=[-\partial^2 + W(r)]\,\psi=\lambda \psi,
\eeq
where $\psi$ generically denotes scalar, fermion or gauge fieds. In
order to perform such a calculation, one should i) choose a suitable
eigenfunction basis; ii) define the renormalization procedure. In
both respects, our approach will be similar to that of
Ref.~\cite{Baacke}, where sphaleron computations within the SM have
been performed. Details will be given in the next subsections; here
we just sketch the main points of our strategy.

Because of the four-dimensional spherical symmetry of the bounce, the
`interaction term' $W$ in~(\ref{eq:eigeq}) depends only on the radial
coordinate $r = (x_\mu x_\mu)^{1/2}$. For this reason it is
convenient to decompose the various fields in eigenstates of the
four-dimensional angular momentum operator $L^2$, and to write the
Laplace operator as
\beq
\label{eq:laplace}
\partial^2=\partial_\mu \partial_\mu=
\frac{d^2}{dr^2}+\frac{3}{r}\frac{d}{dr}-\frac{L^2}{r^2}\equiv\nabla_L~.
\eeq
In the case of scalar fields, the eigenfunctions of $L^2$ are the
four-dimensional spherical harmonics $Y_{j}(\theta)$ 
(where $\theta$ collectively denotes the 3 polar angles)
with eigenvalues
$4j(j+1)$, and degeneracy $(2j+1)^2$, where $j$ takes integer and
semi-integer values (see Appendix~\ref{app:armoniche}). 
After this decomposition, we have
\beq
\label{eq:jsum}
\log\Det S_0''=\sum_j \log\Det(S_0'')_j~,
\eeq
where $(S_0'')_j$ is the restriction of $S_0''$ to the subspace
spanned by eigenfunctions with angular momentum $j$. As we shall see,
the situation is slightly more complicated for fermion and vector
fields. Their expansion in spinor and vectorial hyperspherical
harmonic functions will be obtained starting from the $Y_{j}(\theta)$.

A further simplification arises from the fact that,
in order to compute the ratio in Eq.~(\ref{eq:detratio}),
it is only necessary to solve Eq.~(\ref{eq:eigeq})
for $\lambda=0$. Indeed, we can use the result~\cite{Coleman_T}
\beq
\label{eq:thdet}
\rho_j \equiv 
\frac{\Det\left[-\nabla_L + W(r) \right]}{\Det\left[-\nabla_L \right]} 
=\lim_{r\to\infty} \frac{\det u^j_W(r)}{\det u^j_0(r)}~,
\eeq
where $u^j_W(r)$ [$u^j_0(r)$] are eigenfunctions, regular at $r=0$, 
of $-\nabla_L + W(r)$ [$\nabla_L$] with zero eigenvalues:
\beq
[-\nabla_L + W(r)]\, u^j_W(r)=0,\qquad \nabla_L \, u^j_0(r)=0~.
\eeq
The symbol `det' in Eq.~(\ref{eq:thdet}) stands for the ordinary
determinant over residual (spinorial, gauge group, etc.) indices of
these solutions (see e.g.~\cite{Baacke} and the next subsections for
more details).

\medskip

The one-loop action is affected by the usual ultraviolet divergences
of renormalizable quantum field theories. As a consequence, the sum
over $j$ in Eq.~(\ref{eq:jsum}) is not convergent (the ultraviolet
behaviour being encoded in the behaviour of the determinants for $j\to
\infty$), and the usual renormalization procedure is needed. The
expression
\beq
S_1 = S_0 + \frac{1}{2}\ln \SDet S''_0[\phib]
-\frac{1}{2}\ln\SDet S''_0[0]
\equiv S_0 +\Delta S
\eeq
can be made finite by adding an appropriate set of local counterterms,
which lead to a redefinition of the bare couplings in $S_0$.
For example, in the $\overline{\rm MS}$ scheme (dimensional regularization
with $\overline{\rm MS}$ subtraction) the renormalized one-loop action
can be written as
\beq
\label{subtr}
S_1=S_0^{\overline{\rm MS}} 
+ \left[\Delta S-\left(\Delta S\right)_{\rm pole} \right],
\eeq
where $S_0^{\overline{\rm MS}}$ is the lowest-order action 
expressed in terms of the renormalized $\overline{\rm MS}$ couplings,
and $\left(\Delta S\right)_{\rm pole}$ is the divergent 
part of $\Delta S$ defined according to the $\overline{\rm MS}$
renormalization prescription.
Computing the full determinant in $d=4-2\epsilon$ dimensions 
would be an extremely difficult task. However, this is not necessary.
In fact, the divergent terms are all contained in 
$\Delta S^{[2]}$, defined as 
the expansion of $\Delta S$ up to second order in the `interaction' $W$:
\beqa
\Delta S^{[2]} &=& 
\frac{1}{2}\left[\ln \SDet \left( -\partial^2 + W  \right) -  \ln
\SDet \left( -\partial^2 \right) \right] _{\cO(W^2)} 
\no\\
&=& \frac{1}{2} {\rm STr} \left[ (-\partial^2)^{-1} W \right]
- \frac{1}{4} {\rm STr} \left[ (-\partial^2)^{-1} W (-\partial^2)^{-1} W 
\right]~,
\label{eq:defS2}
\eeqa
where ${\rm STr} ={\rm Tr}$ or ${\rm STr} =- 2{\rm Tr}$ depending on 
whether it acts on boson or fermion fields. In other words, the difference
$\Delta S - \Delta S^{[2]}$ is ultraviolet-finite.
Equation~(\ref{subtr}) can be rewritten as 
\begin{equation}\label{eq:+-+}
S_1= S_0^{\overline{\rm MS}} 
+\bigg[\Delta S-\Delta S^{[2]}\bigg]
+\bigg[\Delta S^{[2]} -\left(\Delta S\right)_{\rm pole}\bigg]~,
\end{equation}
where the two terms in square brackets are separately finite.
The advantage of this last expression is that 
$\Delta S^{[2]}$ can be computed either as a (divergent) sum of terms
corresponding to different values of the angular momentum, which gives
a finite result when subtracted from $\Delta S$, 
or by standard diagrammatic techniques in $4-2\epsilon$ dimensions.
In the next subsections we shall show how
this procedure is implemented in practice.

The final result is expressed in terms of the renormalized parameters
$\lambda,~g_t$ and $g_i$, whose definitions depend on the
renormalization scheme. Of course, the scheme dependence disappears
once these couplings are re-expressed in terms of physical
observables, such as Higgs and top pole masses. In practice, however, in
the case of the gauge couplings it turns out to be more convenient to
directly use the $\overline{\rm MS}$ definitions, since these
parameters are accurately determined by fitting multiple observables.

\subsection{Higgs fluctuations}
The relative corrections to the action due to fluctuations of the Higgs field
are generally small because the Higgs coupling $\lambda$ is small; 
it is however important to consider them,
because they include the special zero-modes (eigenfunctions with zero
eigenvalue) corresponding to translations ($j=1/2$) and field
dilatations ($j=0$) of the bounce, as well as the unique negative
eigenvalue corresponding to space dilatations of the bounce. It is precisely
the existence of this negative eigenvalue that makes
the false vacuum unstable. In this case the interaction term is
simply given by
\beq W(r)=V''(\phib) =-\frac{24 R^2}{(r^2+R^2)^2}~.
\eeq
It does not depend on any coupling constant, as
expected, since the leading contribution to the action is proportional
to $1/\lambda$.

The solutions of the `free equation' $\nabla_L u_0^j(r)=0$, regular in
$r=0$, are immediately found to be proportional to $r^{2j}$
(see Eq.~(\ref{eq:laplace})).
In order to exploit the result in Eq.~(\ref{eq:thdet}),
we must compute the ratio
\beq
\rho_j(r)= \frac{u_W^j(r)}{u_0^j(r)}~.
\eeq
The functions $\rho_j(r)$ obey the differential equations
\beq
\label{eq:Hfl}
\rho''_j(r)+\frac{4j+3}{r} \rho'_j(r) =  W(r)~\rho_j(r)~,
\eeq
which can be solved analytically:
\beq
\rho_j(r) = \frac{R^4 + a_j r^2 R^2 + b_j r^4}{(r^2+R^2)^2}~,
\eeq
where
\beq
\label{eq:ab}
a_j=\frac{2j-1}{j+1};\;\;\;\;b_j=\frac{j(2j-1)}{(j+1)(2j+3)}~.
\eeq
According to Eq.~(\ref{eq:thdet}), we have therefore
\beq
\label{eq:thdethiggs}
\rho_j =
\frac{\Det\left[-\nabla_L + W(r) \right]}{\Det\left[-\nabla_L \right]} 
=\lim_{r\to\infty} \rho_j(r)=b_j~.
\eeq
Let us neglect for the moment the contributions of $j=0$ and $j=1/2$,
for which $\rho_j=0$. Taking into
account the multiplicities of the sub-determinants, we have
\beq
\label{eq:Hserie}
\Delta S^{\rm Higgs}_{j>1/2} = {1 \over 2} \sum_{j>1/2} (2j+1)^2 \ln
\rho_j 
= \sum_{j>1/2} \left[-6j-3-\frac{3}{2j}+\frac{3}{4j^2}+\ldots \right]~.
\eeq 
As discussed in the previous section, 
we regularize this expression by subtracting from it the terms 
obtained by solving Eq.~(\ref{eq:Hfl}) perturbatively 
in $W$. Replacing $W$ with $\epsilon W$, we define 
the coefficients $h_{ij}$ as
$\rho_j(\epsilon) = 1+ \epsilon h_{1j} + \epsilon^2 h_{2j}
+\cO(\epsilon^3)$.
These coefficients can be easily determined numerically. 
The subtracted series is rapidly converging, and in practice the inclusion of
the first ten terms already provides an excellent approximation of the
full result. We find
\beq
\label{eq:Hserie2}
{1 \over 2} \sum_{j>1/2} (2j+1)^2 \ln \rho_j - 
{1 \over 2} \sum_{j\geq 0} (2j+1)^2 \left( h_{1j} - h^2_{1j} + \frac{h_{2j}}{2}
\right)
= 12.6~.
\eeq 

We now turn to a discussion of zero eigenvalues.
The four zero-modes in the $j=1/2$ sector correspond to translations of the
bounce. They can be converted into a volume factor following the
procedure illustrated
in~\cite{coleman_1L}, which amounts to replacing $\rho_{1/2}$ with
\beq
\label{eq:rho12}
\rho'_{1/2}=
\frac{\Det'\left(S_0''[\phib]\right)_{1/2}}{\Det\left(S_0''[0]\right)_{1/2}}
=\lim_{\epsilon\to 0}
\frac{\Det\big[\epsilon+\left(S''_0[\phib]\right)_{1/2}\big]/\epsilon}
{\Det\left(S_0''[0]\right)_{1/2}}~,
\eeq
and multiplying the expression for the tunnelling probability per unit
volume by a factor $\sqrt{S_0[h]/2\pi}$ for each of the four translation
zero modes.
This is the origin of the factor
$S_0^2[\phib]/(2\pi)^2$ in Eq.~(\ref{eq:S1}), which,
in our final result Eq.~(\ref{eq:DeltaS}), is included in $f_h$. 
The elimination of the four vanishing
eigenvalues from $\rho_{1/2}$ provides the dimensional factor $R^{-4}$
in Eq.~(\ref{eq:p1L}). In fact, after numerical 
integration of the corresponding differential equation, we find
\beq
\rho'_{1/2}=0.041 {R^2}~.
\eeq
As can be see from Eq.~(\ref{eq:ab}), there is another zero
eigenvalue in the $j=0$ sector. It arises from the
scale invariance of the tree-level potential: bubbles with different
field value $\sim 1/R$ have the same tree-level action. Scale
invariance is broken by quantum corrections, which shift the zero
eigenvalue by an amount of $\cO[g_i^2/(4\pi R)^2 ]$. We can
therefore compute $\rho_0$ by removing the zero eigenvalue from the
determinant as in Eq.~(\ref{eq:rho12}), and replacing the zero
eigenvalue with a quantity of $\cO[ g_i^2/(4\pi R)^2 ]$ (this small
correction can in principle be extracted from the $R$ dependence of the
one-loop bounce action~\cite{Affleck}). This gives $\rho'_0\sim -R^{2}$, and
\beq
\label{eq:rho0}
{1\over 2} \ln |\rho_0| \sim
{1\over 2} \ln \left|\rho'_0 \frac{g_t^2}{(4\pi R)^2}\right| 
=-2.5 
\eeq
for $\rho'_0=-R^{2}$ and $g_t=1$.
Note that $\rho'_0$ is negative, as it should be because of the
instability of the electroweak vacuum~\cite{coleman_1L}.

\medskip

Combining the results in 
Eqs.~(\ref{eq:Hserie2})--(\ref{eq:rho0}) we finally obtain
\beq
f_h^{(1)}\equiv\left[(\Delta S'-4\ln R)-\Delta S^{[2]} \right]_{\rm Higgs}
=4 \pm 1~,
\label{eq:DSHa}
\eeq
where the quoted error is entirely ascribable to the uncertainty in
Eq.~(\ref{eq:rho0}). In Section~\ref{subs:GG}, $f_h^{(1)}$ will be
combined with analogous contributions generated by Goldstone boson
fluctuations to yield the constant $f_h$ in Eq.~(\ref{eq:DeltaS}).

\subsubsection*{Renormalization}
Following the general strategy discussed in Section~\ref{subs:GS},
we now proceed to evaluate $\Delta S^{[2]}$ in the 
$\overline{\rm MS}$ scheme. The explicit $d$-dimensional
expression of $\Delta S^{[2]}$, defined in
Eq.~(\ref{eq:defS2}), is
\beq
\left[\Delta S^{[2]}\right]^d_{\rm Higgs} =
\frac{3\lambda\mu^{4-d} }{2} ~ \Delta_1 \int d^d x~h^2(x) 
- \frac{9\lambda^2 \mu^{2(4-d)} }{4} \int d^d x ~d^d y~ ~h^2(x) 
\Delta_2(x-y) h^2(y)~,
\label{eq:DSHd}
\eeq
where
\beqa
\Delta_1 &=& \int \frac{d^d k}{(2\pi)^d} \frac{1}{k^2}~, \\
\Delta_2(x-y) &=& \int \frac{d^d q}{(2\pi)^d} e^{iq(x-y)} \int
\frac{d^d k}{(2\pi)^d} \frac{1}{k^2(k+q)^2}~,
\eeqa
and $\mu$ is the usual mass scale that is introduced in dimensional
regularization in order to keep the action dimensionless.
In dimensional regularization, $\Delta_1=0$, and there is thus no
contribution to $\Delta S^{[2]}$ from terms of order $\cO(W)$. 
This is also true for the $\cO(W)$ terms in the fermion and
gauge sectors, as a consequence of our approximation of neglecting
mass terms in the tree-level potential. The second term in
Eq.~(\ref{eq:DSHd}) can be written as
\beq \left[\Delta S^{[2]}\right]^d_{\rm Higgs} = -
\frac{9\lambda^2\mu^{4-d} }{4}
 \int \frac{d^d q}{(2\pi)^d} \left[ \widetilde{h^2}(q^2) \right]^2
B_0(q^2)~,
\label{eq:DSHd1}
\eeq
where $\widetilde{h^2}(q^2)$ is the $d$-dimensional 
Fourier transform of $h^2(x)$ and 
\beq
B_0(q^2) = \mu^{4-d } \int \frac{d^d k}{(2\pi)^d} \frac{1}{k^2(k+q)^2} 
= \frac{1}{(4\pi)^2} \left[ \frac{1}{\bar\epsilon} + 2
+\ln\frac{\mu^2}{q^2}\right]+{\cal O}(\bar\epsilon)~.
\eeq
Here $1/\bar\epsilon=2/(d-4) -\gamma_E + \ln(4\pi)$ denotes 
the divergent part, to be subtracted according to the 
$\overline{\rm MS}$ prescription. We therefore have
\beq
\left(\Delta S\right)_{\rm pole}^{\rm Higgs} = -\frac{1}{\bar\epsilon}\,
\frac{9\lambda^2\mu^{4-d} }{64 \pi^2}
\int \frac{d^d q}{(2\pi)^d} \left[ \widetilde {h^2}(q^2) \right]^2~.
\eeq
Since
\beq 
\int \frac{d^d q}{(2\pi)^d} \left[ \widetilde {h^2}(q^2) \right]^2
= \int d^d x~h^4(x)~,
\eeq
the divergent part in (\ref{eq:DSHd1}) is immediately recognized
to have the same structure as the bare action.

After subtraction of the $1/\bar\epsilon$ pole, the integral in
Eq.~(\ref{eq:DSHd1}) can be safely performed in four dimensions (see
Appendix~\ref{app:fourier}), leading to
\beq
\bigg[\Delta S^{[2]} -\left(\Delta S\right)_{\rm pole}\bigg]_{\rm Higgs}
=-\frac{9\lambda^2}{64\pi^2}
 \int \frac{d^4 q}{(2\pi)^4} 
\left[ \widetilde{h^2}(q^2) \right]^2
\left[ 2+\ln\frac{\mu^2}{q^2}\right]
=-3L -\frac{5}{2}~,
\eeq
where $L=\ln (R\mu e^{\gamma_{\rm E}}/2)$.
Combining the above result with Eq.~(\ref{eq:DSHa}), we 
obtain the full one-loop correction to the action due to fluctuations 
of the Higgs field, renormalized in the $\overline{\rm MS}$ scheme. 

\subsection{Top fluctuations}
The fermionic determinant due to fluctuations of the top-quark 
field assumes a form similar to the bosonic one, if expressed 
in terms of the squared Dirac operator:
\beq
\Dsl^\dagger \Dsl = - \partial^2+\left( \frac{g_t^2}{2} \phib^2
+ \frac{g_t}{\sqrt{2}} \ds \phib \right) \equiv - \partial^2 +W^2~,
\eeq
where $g_t$ is the top Yukawa coupling. Indeed, we can write 
\beq
\Delta S_{\rm top} = -\frac{N_c}{2} \left[ \ln \Det \left( \Dsl^\dagger
\Dsl \right)
- \ln \Det \left(- \partial^2 \right)\right],
\eeq
where $N_c=3$ is the number of colours.

The only difference with respect to the Higgs case is the structure of
$W$. Since $W$ does not commute with $L^2$ because of the 
$\ds\phib$ term, we cannot decompose the eigenfunctions
as products of scalar spherical harmonics times constant Dirac spinors.
However, the spherical symmetry of the bounce, which implies
\beq
\ds \phib(r) = {\hat x \hspace{-1.2ex}/} \phib'(r)~, \qquad {\hat x}_\mu=
x_\mu/r~,
\eeq
considerably simplifies the problem.
Following Ref.~\cite{DeVega}, we decompose the fermionic field as 
\beq
\label{eq:fermion}
\psi (r, \theta) = \sum_{J=\pm 3/2,~\pm 5/2,\ldots } 
\left[
\alpha_{1J}(r) \psi_{1J}(\theta) + \alpha_{2J}(r)\psi_{2J}(\theta)
\right]~,
\eeq
where $\alpha_{iJ}(r)$ are scalar functions and $\psi_{iJ}(\theta)$
are spinors written in terms of appropriate hyperspherical harmonic
functions of multiplicity $J^2 - 1/4$ (see
Appendix~\ref{app:armoniche}). In this basis the equation
$\Dsl^\dagger\Dsl\,\psi=0$ becomes
\beq 
\label{eq:dettop}
\left[\ba{cc}
-\frac{d^2}{dr^2} + \frac{J(J-1)}{r^2} + \frac{g_t^2}{2} \phib^2(r)
                                    &   \frac{g_t}{\sqrt{2}}\phib'(r) 
\\ \frac{g_t}{\sqrt{2}}\phib'(r)
  & -\frac{d^2}{dr^2}   + \frac{J(J+1)}{r^2} +  \frac{g_t^2}{2} \phib^2(r)
 \ea\right] 
\left[ \ba{c} \alpha_{1J}(r) \\ \alpha_{2J}(r) \ea\right] =0~. 
\eeq
In the limit $g_t=0$ the two components of Eq.~(\ref{eq:dettop}) are
decoupled; for $J>0$ their solutions, regular in $r=0$, 
are given by $\alpha^0_{1J} \propto r^{J}$ and $\alpha^0_{2J}\propto
r^{J+1}$. 
Since Eq.~(\ref{eq:dettop}) is invariant under the
exchange $J \leftrightarrow -J$, we shall consider in the following only 
the case $J>0$, and include an extra factor of 2 in the multiplicity
of the solutions.
Equation~(\ref{eq:dettop}) can be cast in the form
\beq
\label{eq:dettop2}
\left\{\begin{array}{rcl}\displaystyle
\rho''_{1J}+2\frac{J}{r}\rho'_{1J} &=& \displaystyle
\frac{g_t^2}{2} \phib^2 \rho_{1J}+ \frac{g_t}{\sqrt{2}}\phib' \rho_{2J}
r~, \\[3mm]
\displaystyle
\rho''_{2J}+2\frac{J+1}{r}\rho'_{2J} &=& \displaystyle
\frac{g_t^2}{2} \phib^2 \rho_{2J}+ \frac{g_t}{\sqrt{2}}\phib'
\rho_{1J} r^{-1}~,
\end{array}\right.
\eeq
where $\rho_{iJ}(r)=\alpha_{iJ}(r)/\alpha^0_{iJ}(r)$.
The system Eq.~(\ref{eq:dettop2}) has 
two solutions $\rho^k_{iJ}(r)$, $k=1,2$
with initial conditions $\rho^k_{iJ}(0)=\delta_{ik}$. 
According to Ref.~\cite{Coleman_T} we can finally write 
\beq
\label{eq:dettop3}
\Delta S_{\rm top}=-3\sum_{J ={\rm I\!N}+1/2}
\left(J^2-\frac{1}{4}\right) \ln\left\{ \lim_{r\to\infty}
\det
\left[\ba{cc} \rho^1_{1J}(r) &
\rho^2_{1J}(r)\cr
\rho^1_{2J}(r) & \rho^2_{2J}(r) \ea\right] \right\}~.
\eeq
The sub-determinants at fixed $J$ 
can be computed by 
means of numerical methods.
As usual, the sum in Eq.~(\ref{eq:dettop3}) is divergent
and we regularize it by subtracting the first two 
powers in $W$. Since 
$g_t$ appears always multiplied by $\phib(r)$,
the final result depends only on the ratio $g^2_t/|\lambda|$:
\beq
\bigg[\Delta S-\Delta S^{[2]}\bigg]_{\rm top}
= -f_t\left(\frac{g^2_t}{|\lambda|} \right)~,
\eeq
and is plotted in Fig.~\ref{fig:ft} in the range of interest.

The evaluation of $\Delta S^{[2]}$ in dimensional regularization 
is very similar to the Higgs case. We find 
\beq
\left[\Delta S^{[2]}\right]^d_{\rm top} = 
\frac{N_c}{4} \int \frac{d^d q}{(2\pi)^d} 
 \left\{ g_t^4 \mu^{4-d}[ \widetilde{h^2}(q^2)]^2 
+ 2 g_t^2 q^2 {\tilde h}^2(q^2) \right\} B_0(q^2)~,
\eeq
which, after subtraction of the divergent terms, leads to 
\beq
\bigg[\Delta S^{[2]} -\left(\Delta S\right)_{\rm pole}\bigg]_{\rm top}
= \frac{g_t^4}{6|\lambda|^2}\left( 5 + 6L\right)
+\frac{g_t^2}{6|\lambda|}\left( 13 + 12L\right)~.
\eeq

\subsection{Gauge and Goldstone fluctuations}
\label{subs:GG}
In order to evaluate quantum fluctuations in the sector of gauge
fields we need to modify the bare action introducing an appropriate
gauge fixing. Both the correction to the potential and that to the kinetic
term are gauge-dependent, but they combine to give gauge-independent
physical quantities~\cite{Frere}. For example, at one loop, the
divergent corrections change the Higgs action into
\beq
S = \int d^4 x \left[(1+Z_T) |D_\mu \phi|^2 + (1+Z_V) \lambda |\phi|^4\right]~,
\eeq
and the gauge dependence of $Z_T$ and $Z_V$ cancels in the combination
$Z_V-2 Z_T$ that determines the RGE equation for $\lambda$. By
requiring that the bounce action be an extremum under the
transformation $h(r)\to h(\xi r)$, one finds that $S_T = -2 S_V$,
where $S_T$ ($S_V$) is the `kinetic part' (`potential part') of the
tree-level bounce action $S=S_T+S_V$. Therefore, the divergent
corrections to the bounce action are gauge-independent. More
generally, it was shown by Nielsen~\cite{Nielsen} that the value of
the effective potential at its minimum is gauge-invariant. Similarly,
the value of the effective action is also invariant, when evaluated
over a field configuration that extremizes it. It follows that the
full corrections to the bounce action are gauge-independent.

The functional obtained after gauge fixing depends on gauge fields
$A^a_\mu$, Goldstone bosons $G^b$ and Faddeev--Popov ghosts $\eta^c$.
We are only interested in the second functional derivative of the
action evaluated at $A^a_\mu=G^b=\eta^c=0$ and $h\not=0$. This
considerably simplifies the problem, since we can ignore the
non-Abelian part of gauge interactions. In the following, we will denote
with a suffix $A$ the corrections to the action for a generic abelian
gauge field $A_\mu$, with coupling $g$.
We can also ignore the
fluctuations of the electromagnetic field: they vanish in the
difference $\Delta S[h]-\Delta S[0]$ because the photon field is not
coupled to $h$. The problem is thus equivalent to the case of three
independent Abelian fields: the two $W$'s and the $Z$.

Introducing a 't Hooft--Feynman gauge-fixing term, which
has proved to be the most convenient choice for our 
calculation, the relevant part of the bare action 
can be written as
\beq
\label{eq:g_action}
\int d^4 x~ \left[\frac{1}{4} F_{\mu\nu}^2 
+ \frac{1}{2} (\partial A- g h G)^2 + |D\phi|^2 + V(\phi) +
\eta^*[ -\partial^2 + M(G,h) ] \eta\right]~,
\eeq
where $M(0,h)=g^2 h^2/4$. This choice
of the gauge fixing term has the advantage that all terms of the type
$A_\mu\partial_\mu G$ and $\partial_\mu\partial_\nu A_\mu$, generated
by $|D\phi|^2$ and $F_{\mu\nu}^2$, respectively, are eliminated from
the equations of motion.

The ghost fields can be treated separately since the the second 
derivative of the action is diagonal with respect to them. 
On the contrary, we cannot separate $A_\mu$ and $G$, 
whose zero-eigenvalue equation is given by 
\beq
\left[\ba{cc} -\partial^2\delta_{\mu\nu}
+\frac{1}{4} g^2 h^2(r)\delta_{\mu\nu}
& -g \hat{x}_\mu h'(r)  \\
-g \hat{x}_\nu h'(r)  & -\partial^2
+\left(\frac{1}{4}g^2 + \lambda\right) h^2(r)
\ea \right] \left[ \ba{c} A_\nu(x) \\ G(x) \ea\right] = 0~.
\label{eq:gauge1}
\eeq
In order to solve Eq.~(\ref{eq:gauge1}) we need a suitable 
decomposition of $A_\mu$ in terms of spherical harmonics. 
A convenient choice is provided by 
\beq
\label{eq:amu}
A_\mu(x) = \sum_{j=0,\frac{1}{2},1,\ldots} \bigg[a_{1j}(r) \hat{x}_\mu
+ \frac{a_{2j}(r) }{2 [j(j+1)]^{1/2}} 
{r\partial_\mu} + 
\sum_{i=3,4} a_{ij}(r) \epsilon_{\mu\nu\rho\sigma} V_\nu^{(i)} 
x_\rho \partial_\sigma\bigg] Y_j(\theta)~,
\eeq
where the $a_{ij}(r)$ are scalar functions and $V_\nu^{(i)}$ are two
generic orthogonal vectors. In this basis the operator $L^2$ is
diagonal with respect to the index $j$ and, at fixed $j$, mixes only
the first two components, $a_{1j}$ and $a_{2j}$.
Since $\partial_\mu Y_0 =0$, the case $j=0$ deserves special attention.
For this reason, we discuss separately the two cases $j>0$ and $j=0$.

\subsubsection*{Sub-determinants with $j>0$}
As shown in
Appendix~\ref{app:armoniche}, for $j\not=0$ the zero-eigenvalue
equation is
\beq
\label{eq:gauge2}
\left[\ba{ccc}
-\nabla_{L}+\frac{3}{r^2} +\frac{1}{4} g^2 h^2(r)& 
-\frac{4\sqrt{j(j+1)}}{r^2}  &
-gh'(r) \cr
-\frac{4\sqrt{j(j+1)}}{r^2}  & 
\!\!\! -\nabla_{L}-\frac{1}{r^2} +\frac{1}{4} g^2 h^2(r) & 
0\cr
-gh'(r)  & 
0  
& \!\!\! -\nabla_{L}+(\frac{1}{4}g^2+\lambda) h^2(r) \ea\right]
\left[ \ba{c} a_{1j}(r)  \cr a_{2j}(r) \cr G_j(r) \ea\right] = 0~, 
\eeq
\beq
\left[ -\nabla_{L} +\frac{1}{4} g^2 h^2(r) \right]  a_{ij}(r) =0~, 
\;\;\;\;i=3,4~,
\eeq
where $G_j(r)$ are the components of a standard
decomposition of $G(x)$ in spherical harmonics.

Interestingly, the components $a_{3j}$ and $a_{4j}$, which are
decoupled from the Goldstone boson sector, obey exactly the same
equation as the two components of the complex ghost field. The latter
contribute to the action with an opposite sign; therefore these two
contributions cancel against each other for $j\not=0$.

Equation~(\ref{eq:gauge2}) can be further simplified by means of 
a rotation in the $(a_{1j},a_{2j})$ space,
\beq
\left[ \ba{c} a_{1j} \\ a_{2j} \ea\right]  =
\frac{1}{\sqrt{2j+1}}
\left[ \ba{cc}  \sqrt{j}  & - \sqrt{j+1} \cr  
               \sqrt{j+1} &  \sqrt{j}
\ea\right]
\left[ \ba{c} {\tilde a}_{1j} \\ {\tilde a}_{2j} \ea\right]~,\qquad
\eeq
which diagonalizes it in the gauge sector. In this new basis we have 
\beq
\left[\ba{ccc}
-\nabla_{L_-} +\frac{1}{4} g^2 h^2 &0  &
-gh'\sqrt{j/(2j+1)} \cr
0  & -\nabla_{L_+} +\frac{1}{4} g^2 h^2 & gh' \sqrt{(j+1)/(2j+1)} \cr
  -gh'\sqrt{j/(2j+1)}     &  gh' \sqrt{(j+1)/(2j+1)}  &
-\nabla_{L}+( \frac{1}{4}g^2+\lambda) h^2 \ea \right]
\left[ \ba{c} {\tilde a}_{1j}  \cr {\tilde a}_{2j} \cr G_j \ea\right] = 0~, 
\eeq
where $\nabla_{L_\pm}$ denote the usual $\partial^2$ operator
at fixed angular momentum, with eigenvalues shited 
by $j\to j\pm1/2$. The 
free equation ($h\to 0$) is now diagonal and, similarly to the
top and Higgs cases, can be easily solved.
We finally obtain
\beq
\left\{\begin{array}{rcl}
\displaystyle\rho''_{1j} + \frac{1+4j}{r} \rho'_{1j}
&=&\displaystyle  \frac{1}{4} g^2 h^2 \rho_{1j} 
- r\rho_{3j}  gh' \sqrt{\frac{j}{2j+1}} \\[2mm]
\displaystyle\rho''_{2j} + \frac{5+4j}{r} \rho'_{2j}  
&=&\displaystyle  \frac{1}{4} g^2 h^2 \rho_{2j} 
+ \frac{1}{r}\rho_{3j}gh' \sqrt{\frac{j+1}{2j+1}} \\[2mm]
\displaystyle\rho''_{3j} + \frac{3+4j}{r} \rho'_{3j}  
     &=&\displaystyle \left(\frac{1}{4} g^2 h^2 +\lambda h^2\right)\rho_{3j}  
- \frac{1}{r}\rho_{1j} gh' \sqrt{\frac{j}{2j+1}}
+ r\rho_{2j} gh' \sqrt{\frac{j+1}{2j+1}}
\end{array}\right.~,
\eeq 
where, as in the previous cases, we have defined
\beq
\rho_{ij}=\frac{{\tilde a}_{ij}}{{\tilde a}_{ij}^0},\;\;i=1,2;\;\;\;\;
\rho_{3j}=\frac{G_j}{G_j^0}~.
\eeq
In complete analogy with the top case, the above system has 
three solutions $\rho^k_{ij}(r)$ ($k=1,2,3$),
with initial conditions  $\rho^k_{ij}(0)=\delta_{ik}$, 
and the resulting correction to the action can be written as 
\beq
\label{eq:gauge3}
\Delta S^A_{j>0} = \frac{1}{2} \sum_{j>0}
(2j+1)^2 \ln\left\{ \lim_{r\to\infty}
\det
\left[\ba{ccc} \rho^1_{1j}(r) & \rho^2_{1j}(r) & \rho^3_{1j}(r)\cr
\rho^1_{2j}(r) & \rho^2_{2j}(r) & \rho^3_{2j}(r) \cr
\rho^1_{3j}(r) & \rho^2_{3j}(r) & \rho^3_{3j}(r) \ea\right] \right\}~.
\eeq
All the determinants in Eq.~(\ref{eq:gauge3}) 
are different from zero and, as usual, their sum must be regularized 
by the subtraction of $\Delta S^{[2]}$.

\subsubsection*{Sub-determinant with $j=0$}
For  $j=0$  only 
the $\hat{x}_\mu$ component of $A_\mu$ 
is different from zero 
[$A_\mu^{(j=0)} = a_{10}(r) \hat{x}_\mu$] 
and the eigenvalue equation is
\beq
\label{eq:gauge01}
\left[\ba{cc}
-\frac{d^2}{dr^2}-\frac{3}{r}\frac{d}{dr}
+\frac{3}{r^2} +\frac{1}{4} g^2 h^2(r) & 
-gh'(r) \cr -gh'(r) &
-\frac{d^2}{dr^2}-\frac{3}{r}\frac{d}{dr}
+(\frac{1}{4}g^2+\lambda) h^2(r) \ea\right]
\left[ \ba{c} a_{10}(r) \cr G_0(r) \ea\right] = 0~.
\eeq
This leads to
\beq
\label{eq:gauge02}
\left\{\begin{array}{rcl}
\displaystyle\rho''_1 + \frac{5}{r} \rho'_1
&=&\displaystyle  \frac{1}{4} g^2 h^2 \rho_1 
- \frac{1}{r}\rho_2 gh'  \\[2mm]
\displaystyle\rho''_2 + \frac{3}{r} \rho'_2
&=&\displaystyle \left(\frac{1}{4} g^2 h^2 +\lambda h^2\right)\rho_2
- r\rho_{1} gh' 
\end{array}\right.
\eeq 
where, as usual,  $\rho_{1}=a_{10}/a_{10}^0$ and $\rho_{2}=G_0/G_0^0$. 
The absence of transverse components in $A^{(j=0)}_\mu$
implies that the $j=0$ sub-determinant of the ghost fields
is not cancelled; it is obtained as the large-$r$ limit 
of the function $\rho_\eta(r)$, where
\beq
\rho_\eta'' + \frac{3}{r} \rho'_\eta
=  \frac{1}{4} g^2 h^2 \rho_\eta~.
\eeq
Employing the usual notation for the two solutions of 
the system (\ref{eq:gauge01}), we can write
\beq
\label{eq:gauge03}
\Delta S^A_{j=0} = \frac{1}{2} \ln\left\{ \lim_{r\to\infty}
\det
\left[\ba{cc} \rho^1_{1}(r) & \rho^2_{1}(r)  \cr
\rho^1_{2}(r) & \rho^2_{2}(r) \ea\right] \right\}
- \ln  \lim_{r\to\infty} \rho_\eta(r) ~.
\eeq
The $2\times 2$ determinant in Eq.~(\ref{eq:gauge03}) contains a
vanishing eigenvalue. This is present also in the limit $g=0$, and
originates from the global symmetry of the action, which persists even
after the introduction of the gauge-fixing term. This corresponds to
the possibility, for the false vacuum, to tunnel in different directions
with equal probabilities. As discussed in
\cite{Baacke,Kusenko}, the fluctuations corresponding to global
rotations can be converted into an integral over the group volume.
These zero-mode fluctuations probe the non-abelian structure of the
gauge group, so that we can no longer compute separately the
$Z,W^\pm$ contributions. Considering the full gauge group, we obtain
\beq \label{eq:IR_h2}
e^{-\Delta S_{\rm gauge}} =  16\pi^2 
\bigg[\frac{1}{2\pi}\int d^4 x ~h^2(r)\bigg]^{3/2}e^{- \Delta S'_{\rm gauge}}
\equiv R^3 e^{-f_h^{(2)}- \Delta S'_{\rm gauge}}~,
\eeq
where the factor $16\pi^2$ is the volume of the broken $SU(2)$ group.
The resulting integral of $h^2$ is logarithmically infrared
divergent, as a consequence of our
approximation of neglecting the mass term in the Higgs potential, which
would have acted as an infrared regulator.  For this reason, we cut off
the integral in Eq.~(\ref{eq:IR_h2}) at $r=1/v$. The result, for
$Rv\ll 1$, is
\beq
\label{eq:f_h2}
f_h^{(2)} =
-\frac{3}{2}\ln\left[\frac{2^{17/3}\pi^{7/3}}{|\lambda|}
\ln \frac{1}{Rv}
\right] + {\cal O}(1/\ln Rv)~.
\eeq
This vanishing eigenvalue is the only aspect of our calculation that
is sensitive to the infrared behaviour of the potential, and therefore
beyond the control of our approximation. However, the sensitivity to
the infrared cut-off is mild and does not induce an appreciable
numerical uncertainty.

\begin{figure}[t]
\begin{center}
\includegraphics{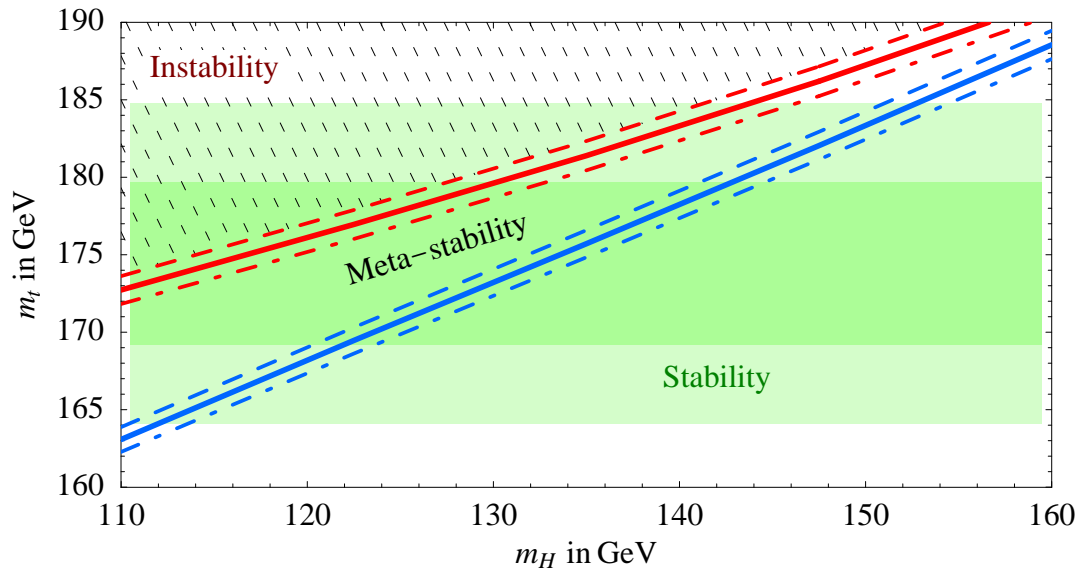}
\caption{\em Metastability region of the
Standard Model vacuum in the $(m_H,m_t)$ plane, for
$\alpha_s(m_Z)=0.118$ (solid curves). Dashed and dot-dashed curves
are obtained for $\alpha_s(m_Z)=0.118\pm 0.002$. 
The shaded area indicates the experimental range for $m_t$. 
Sub-leading effects could shift the bounds by $\pm 2\GeV$ in $m_t$.}
\label{fig:inout}
\end{center}
\end{figure}

Once the vanishing eigenvalues have been removed from the $j=0$
sub-determinant, the latter acquires a dimensional factor that is
compensated by the factor $R^3$ in Eq.~(\ref{eq:IR_h2}).

\subsubsection*{Final result for the gauge sector}
Summing the regularized gauge and Goldstone corrections, we finally obtain
\beq
\Delta S_{\rm gauge} =  2
f_g\left(\frac{g_2^2}{|\lambda|}\right) +
f_g\left(\frac{g_Z^2}{|\lambda|}\right) + f_h^{(2)}+ 3f_h^{(3)}~,
\eeq
where
\beq
f_h^{(3)}=\lim_{g\to0} \left[ (\Delta S'_{j=0} -\ln R) +
\Delta S_{j>0}-\Delta S^{[2]} \right]^A\approx 2
\eeq
and
\beq
f_g\Big(\frac{g^2}{|\lambda|}\Big)
=\left[ (\Delta S'_{j=0} -\ln R) +
\Delta S_{j>0}-\Delta S^{[2]} \right]^A-f_h^{(3)}~,
\eeq
so that $f_g(0)=0$.  The function $f_g$, obtained by means of
numerical integration, is plotted in Fig.~\ref{fig:ft}.  The constants
$f_h^{(2)},f_h^{(3)}$ can finally be combined with $f_h^{(1)}$ in
Eq.~(\ref{eq:DSHa}) and with the translation prefactors
$S_0^2/(2\pi)^2$ to yield
\beq
\label{eq:f_h}
f_h = f_h^{(1)} +f_h^{(2)}+ 3f_h^{(3)} - \ln \frac{16\pi^2}{9\lambda^2}=
 \frac{7}{2}\ln |\lambda| - 8\pm 1~.
\eeq
This numerical estimate has been obtained with $Rv = 10^{-14}$, but it
is very mildly sensitive to the value of $Rv$.

In analogy with the previous cases, dimensional regularization leads to
\beq
\left[\Delta S^{[2]}\right]^d_A = 
-\frac{1}{4}\int \frac{d^d q}{(2\pi)^d}
\bigg\{\bigg[(d-2)\frac{g^4}{16}
+\left(\frac{g^2}{4}  + \lambda\right)^2 \bigg]\mu^{4-d}
[ \widetilde{h^2}(q^2)]^2 - 2 g^2  q^2  {\tilde h}^2(q^2)\bigg\}B_0(q^2)~,
\eeq
which, after subtraction of the divergent terms, reduces to
\beq
\bigg[\Delta S^{[2]} -\left(\Delta S\right)_{\rm pole}\bigg]^A
=-\frac{5+6L}{18}
-\frac{g^2}{|\lambda|}\frac{7+6L}{12}-\frac{g^4}{\lambda^2}\frac{1+2L}{32}.
\eeq

\mysection{Results}
Using Eq.~(\ref{eq:p1L}), the condition $p<p_{\rm max}$ can be 
translated into an upper bound on $|\lambda(\mu)|$:
\beqa
|\lambda(2 e^{-\gamma_E}/R )| &<& 
\frac{8\pi^2/3}{\ln(V_U/R^4)-\Delta S -\ln p_{\rm max}} \no \\
&=& \frac{0.065}{1-\left[\ln(Rv)+\ln(T_U^0/T_U)\right]/100 - 
\left[\Delta S +\ln p_{\rm max}\right]/400}~, \qquad
\label{eq:finale}
\eeqa
where $\Delta S$ is computed at $L=\ln(R\mu e^{\gamma_E}/2)=0$,
$V_U=T_U^4$ and $T_U^0 =10^{10}~{\rm yr}$. Equation~(\ref{eq:finale})
must be satisfied for all values of $R$. From this inequality, a lower
limit on $m_H$ can be obtained by integrating the two-loop RGE
equations~\cite{Ford} for $\lambda$, $g_t$ and the three gauge
couplings $g_i$.  The initial values of $\lambda$ and $g_t$ at $\mu=v$
are related to the values of $m_H$ and $m_t$ by the matching
conditions given in~\cite{SZ} and~\cite{HK}, respectively. As in most
recent analyses of the Higgs potential~\cite{stab_new}, we include the
finite two-loop QCD correction in the relation between $g_t(m_t)$ and
the top pole mass $m_t$. The latter is formally a higher-order
contribution, and can be used to estimate the theoretical uncertainty
in the determination of $\lambda(\mu)$ at high scales.

Stability and metastability bounds in the $(m_H,m_t)$ plane are shown
in Fig.~\ref{fig:inout}, which has been obtained with $p_{\rm max}=1$.
The metastability condition can be approximated as
\beq
\label{eq:mstabbound}
m_H({\rm GeV})
> 117 + 2.9\left[m_t({\rm GeV}) -(175\pm 2)\right]
-2.5 \left[\frac{\alpha_s(m_Z)-0.118}{0.002}\right]
+0.1 \ln\left(\frac{ T_U }{10^{10}~{\rm yr}}\right)~,
\eeq
while the absolute stability bound is given by
\beq
\label{eq:stabbound}
m_H({\rm GeV}) > 133 + 2.0\left[m_t({\rm GeV}) -(175\pm 2)\right]
-1.6 \left[\frac{\alpha_s(m_Z)-0.118}{0.002}\right]~.
\eeq
The numerical impact of $\Delta S$ is rather weak, provided the
renormalization scale is chosen such that $L=0$: for $m_H=115$~GeV, a
correction $\Delta S = 50$ shifts the metastability bound by $1\GeV$
in $m_t$.  For this reason, our results are numerically close to those
obtained in~\cite{QE} (at zero temperature), where this correction was
not included.  Note that one could have expected a larger effect,
because of the large value of the top Yukawa coupling $g_t$ at the
weak scale.  However, the relevant quantity here is $g_t(1/R)$, with
$1/R\sim 10^{16}$~GeV, which is substantially smaller than $g_t(m_t)$.

The uncertainties in the determination of $\Delta S$ turn out to be
negligible with respect to the uncertainties involved in the determination
of $\lambda(\mu)$ at high scales.  We estimate the latter to generate
an error $\delta m_t \sim 2$~GeV at fixed $m_H$ and $\alpha_s$, both in
the stability and metastability bounds, as shown in
Eqs.~(\ref{eq:mstabbound}) and~(\ref{eq:stabbound}).

Some comments are in order:
\begin{itemize}
  
\item Vacuum decay can also be catalysed by collisions of cosmic rays.
  However, it has been shown~\cite{cosmici} that the tunnelling rate
  induced by such processes is negligible with respect to the probability
  of quantum tunnelling.

\item Thermal tunnelling gives stronger bounds than quantum
  tunnelling only under the assumption that the temperature of the universe
  has been above $10^8-10^9$~GeV (for $m_H=115$~GeV).
  
\item We have assumed the validity of the Standard Model up to the
  Planck scale. Experiments at Tevatron run IIB can reduce the error
  on $m_t$ down to $\pm 2\GeV$~\cite{runII} and possibly discover the
  Higgs and measure its mass if $m_H < 130$~GeV. Some extension of the
  SM will become necessary, should $m_H$ and $m_t$ be found in the
  excluded region.  All concrete modifications invoked to cure this
  problem give a computable correction to the squared Higgs mass; for
  example, a scalar with mass $\tilde{m}$ and a coupling
  $\tilde{\lambda}$ to the Higgs gives $\delta m_H^2 \sim
  \tilde{\lambda}\tilde{m}^2/(4\pi)^2$. Since $\tilde{\lambda}$
  cannot be too small, by naturalness arguments we expect the scale
  $\tilde{m}$ of new physics to be in the electroweak range.
  
\item All observed neutrino anomalies could be explained in terms of
  neutrino oscillations without affecting our result~\cite{Vissani}.

\end{itemize}

\mysection{Conclusions}
All recent LEP1, LEP2, SLD and Tevatron data are compatible with the
Standard Model with $m_t\approx 175\GeV$ and a light Higgs, maybe
$m_H\approx 115\GeV$~\cite{Lep2_ev}. It is well known that the
Standard Model is affected by the so-called hierarchy problem, which
manifests itself through uncomputable quadratically divergent
corrections to the Higgs squared mass. The extensions of the theory
proposed to cure this potential problem are likely to involve the
presence of new phenomena not far above the electroweak scale.
However, a definite solution of the naturalness problem is not yet
established, and it is interesting to consider the possibility that it
be solved by some unknown mechanism that takes place around the Planck
scale.  Even within this assumption, a possible phenomenological
problem affects the SM, namely the possibility that the scalar
potential become unbounded from below at large field values, below the
Planck scale.  For $m_H=115\GeV$ such instability is present if the
top mass is larger than $(166\pm 2)\GeV$, i.e.\ only about two
standard deviations below the central value of the Tevatron data.

The instability of the SM vacuum does not contradict any experimental
observation, provided its lifetime is longer than the age of the universe.
In semi-classical approximation, this happens if the running quartic
Higgs coupling $\lambda$ is larger than about $-0.05$. In this paper
we have presented a more accurate assessment of this bound by
performing a full next-to-leading order computation of the tunnelling
rate of the metastable vacuum.  In particular, we have computed the
one-loop corrections to the bounce action.  Our result is summarized
in Eq.~(\ref{eq:risultato}) and plotted in Fig.~\ref{fig:inout}.
Next-to-leading corrections turn out to be numerically small, with an
appropriate choice of the renormalization scale ($\mu=1/R$), but they
allow fixing the various ambiguities of the semi-classical calculation,
thus reducing the overall uncertainty of the result.  We find that,
for $m_H=115\GeV$ (that is, just above the present exclusion limit),
the instability is dangerous only for $m_t > (175\pm 2)\GeV$; this
result does not depend on the unknown new phyiscs at the Planck scale.
A determination of $m_t$ at this level of accuracy would therefore be
extremely interesting in this respect, and will probably be achieved
at Tevatron run II.

\section*{Acknowledgements}
We thank C.~Becchi, P.~Menotti, R.~Rattazzi and M.~Zamaklar for
discussions and suggestions.

\appendix

\mysection{Spherical harmonics}
\label{app:armoniche}
The spherical harmonics $Y_{jmm'}(\theta)$ are eigenfunctions of 
$L^2=L_{\mu\nu}L_{\mu\nu}$, where
\beq
\label{eq:Lmunu}
L_{\mu\nu}=\frac{i}{\sqrt{2}}\left(x_\mu\partial_\nu-x_\nu\partial_\mu\right)~,
\eeq
with eigenvalues $4j(j+1)$; the indices $m,m'$ range between $-j$ and
$j$ (with $j=0,1/2,1,\ldots$) giving the multiplicity $(2j+1)^2$.
This multiplicity can be understood by noting that the group of rotations
in four dimensions is $O(4)\sim SU(2)\otimes SU(2)$, and its scalar
representations correspond to $SU(2)\otimes SU(2)$ representations
$(l,k)$ with $l=k$~\cite{matematici}. The indices $m,m'$ will be
omitted in the following.

Spin-$1/2$ representations are obtained for $l=k\pm 1/2$.  The index
$J$ that appears in Eq.~(\ref{eq:fermion}) is related to $l$ and $k$
through
\beqa
J=2k+\frac{3}{2} \qquad &&{\rm for}\qquad l=k+\frac{1}{2},\; k=0,1/2,1,\ldots~;
\\
J=-2k-\frac{1}{2}\qquad &&{\rm for}\qquad l=k-\frac{1}{2},\; k=1/2,1,\ldots~.
\eeqa
Therefore, the multiplicity of the representation labelled by $J$ is
$(2k+1)(2l+1)=J^2-1/4$. The action of the operator $\Dsl$ on the
$(l,k)$ spinor representations can be found in Ref.~\cite{DeVega}.
                                
In the case of vector representations, it is straightforward to work out the
action of $\partial^2$ on $A_\mu(x)$ as in Eq.~(\ref{eq:amu}).
Using Eq.~(\ref{eq:laplace}) we find
\beqa
\partial^2 a_{1j} \hat{x}_\mu Y_j(\theta) 
&=&
\bigg[a''_{1j}+3\frac{a'_{1j}}{r}-a_{1j}
\frac{L^2+3}{r^2}\bigg]\hat{x}_\mu Y_j(\theta)
+ a_{1j}\frac{2}{r} \partial_\mu Y_j(\theta)
\\
\partial^2 a_{2j} r\partial_\mu Y_j(\theta) 
&=&\bigg[a''_{2j} + 3\frac{a'_{2j}}{r} - a_{2j} \frac{L^2-1}{r^2}
\bigg]r\partial_\mu Y_j(\theta)
+a_{2j} \frac{2L^2}{r^2}\hat{x}_\mu Y_j(\theta)\qquad
\\
\partial^2 a_{ij}\epsilon_{\mu\nu\rho\sigma} V_\nu^{(i)} 
x_\rho \partial_\sigma Y_j(\theta) 
&=&\bigg[a''_{ij} + 3\frac{a'_{ij}}{r} - a_{ij} \frac{L^2}{r^2}
\bigg] \epsilon_{\mu\nu\rho\sigma} V_\nu^{(i)} 
x_\rho \partial_\sigma Y_j(\theta)~.
\eeqa
Therefore, we have
\beq
\int d^4 x A_\mu \partial^2 A_\mu=
\sum_j\int r^3 dr~a_{ij}(r)D^{(j)}_{ik}a_{kj}(r)~,
\eeq
where, with an appropriate normalization of the two vectors $V_\mu^{(i)}$,
\beq
D^{(j)}=\frac{d^2}{dr^2}+\frac{3}{r}\frac{d}{dr}-\frac{4j(j+1)}{r^2}
+\left[\ba{cccc}
-\frac{3}{r^2} & \frac{4\sqrt{j(j+1)}}{r^2} & 0 & 0\cr
 \frac{4\sqrt{j(j+1)}}{r^2} & \frac{1}{r^2} & 0 & 0\cr
     0         &          0    & 0 & 0\cr 
     0         &          0    & 0 & 0\ea\right]~.
\eeq

\mysection{Fourier transform of the bounce}
\label{app:fourier}
The four-dimensional Fourier transform of a function $f$, depending
only on the radial coordinate $r$, is given by
\beq
{\tilde f}(p^2)\equiv\int d^4x e^{i p_\mu x_\mu} f(r) = \frac{4\pi^2}{p}
\int_0^\infty dr\,r^2 f(r) J_1(pr)~,
\eeq
where $p^2=p_\mu p_\mu$ and
\beq
J_1(pr) = 
\frac{pr}{\pi} \int_{0}^\pi d\theta\, \sin^{2}\theta\, e^{ip r\cos\theta}~.
\eeq
In the case of the bounce we find 
\beqa
{\tilde h}(p^2) &=& \frac{4\pi^2}{p}\int_0^\infty dr\,r^2   h(r) J_1(pr)~
=~\frac{8\sqrt{2}\pi^2 R^2 K_1(pR)}{p \sqrt{|\lambda|}}~, \\
\widetilde{h^2}(p^2) &=& \frac{4\pi^2}{p}
\int_0^\infty dr\,r^2   h^2(r) J_1(pr)~
=~\frac{16\pi^2 R^2}{|\lambda| } K_0(p R)~.
\eeqa
The functions $K_n(x)$ and $J_n(x)$ are Bessel functions defined as in
Mathematica~\cite{Mathematica}.


\newpage
\footnotesize
\begin{thebibliography}{999}

\bibitem{stab}
N.~Cabibbo, L.~Maiani, G.~Parisi and R.~Petronzio, \npb{158}{1979}{295};
P.Q.~Hung, \prl{42}{1979}{873};
M.~Lindner, \zpc{31}{1986}{295};
M.~Lindner, M.~Sher and H.~Zaglauer, \plb{228}{1989}{139}.

\bibitem{sher}
M.~Sher, \prep{179}{1989}{273};
B.~Schrempp and M.~Wimmer, \ppnp{37}{1996}{1}.

\bibitem{stab_new}
M.~Sher, \plb{317}{1993}{159}; {\em addendum}~\ibid{331}{1994}{448};
G.~Altarelli and G.~Isidori, \plb{337}{1994}{141};
J.A.~Casas, J.R.~Espinosa and M.~Quir\'os, \plb{342}{1995}{171};
T.~Hambye and K.~Riesselmann, \prd{55}{1997}{7255}.
\bibitem{LEPbound}
P.~Abreu {\it et al.} [DELPHI Collaboration], \plb{499}{2001}{23};
G.~Abbiendi {\it et al.} [OPAL Collaboration], \plb{499}{2001}{38};
P.~Igo-Kemenes, LEPC presentation, Nov. 3, 2000,
{\tt http://lephiggs.web.cern.ch/LEPHIGGS/talks/index.html}.

\bibitem{Lep2_ev} 
R.~Barate {\it et al.}  [ALEPH Collaboration],
\plb{495}{2000}{1};
M.~Acciarri {\it et al.}  [L3 Collaboration],
\plb{495}{2000}{18}.

\bibitem{Higgs@Lep} 
LEP Electroweak Working Group,
{\tt http://www.web.cern.ch/LEPEWWG}.

\bibitem{HBNielsen}
The case $\lambda(\Mp)=0$ has been considered in 
D.L.~Bennett, H.B.~Nielsen and I.~Picek, \plb{208}{1988}{275};
C.D.~Froggatt and H.B.~Nielsen, \plb{368}{1996}{96}.

\bibitem{coleman_sc} 
S.~Coleman, \prd{15}{1977}{2929}.

\bibitem{LeeWeinberg}
K.~Lee and  E.J.~Weinberg, \npb{267}{1986}{181}.

\bibitem{coleman_1L}
C.G.~Callan and S.~Coleman, \prd{16}{1762}{1977}.

\bibitem{Langer}
J.~Langer, {\it Ann. Phys.} {\bf 41} (1967) 108 ;
{\it ibid.} {\bf 54} (1969) 258 ;
{\it Physica} {\bf 73} (1974) 61 .

\bibitem{QE}
J.R.~Espinosa and M.~Quiros,
\plb{353}{1995}{257}.

\bibitem{Ford}
C.~Ford, D.R.~Jones, P.W.~Stephenson and M.B.~Einhorn, \npb{395}{1993}{17}.

\bibitem{SZ}
A.~Sirlin and R.~Zucchini, \npb{266}{1986}{389}.

\bibitem{HK}
R.~Hempfling and B.A.~Kniehl, \prd{51}{1995}{1386}.

\bibitem{Baacke}
J.~Baacke and S.~Junker, \prd{49}{1994}{2055};
\ibid{50}{1994}{4227}.

\bibitem{Coleman_T}
S.~Coleman, ``The uses of instantons'',
{\it Lecture delivered at 1977 Int. 
School of Subnuclear Physics, Erice, Italy, 1977};
R.~Dashen, B.~Hasslacher and A.~Neveu, \prd{10}{1974}{4114}.

\bibitem{Affleck}
I.~Affleck, \npb{191}{1981}{429}.

\bibitem{DeVega}
J.~Avan and H.J.~De Vega, \npb{269}{1986}{621}.

\bibitem{Frere}
J.-M.~Fr\`ere and P.~Nicoletopoulos, \prd{11}{1975}{2332}.

\bibitem{Nielsen}
N.K.~Nielsen, \npb{101}{1975}{173}.

\bibitem{Kusenko}
A.~Kusenko, K.~Lee and E.J.~Weinberg, \prd{55}{1997}{4903}.

\bibitem{cosmici}
K.~Enqvist and J.~McDonald, \npb{513}{1998}{661}. 

\bibitem{Vissani}
F.~Vissani, \prd{57}{1998}{7027}.

\bibitem{runII}
G. Brooijmans (for the CDF and D0 collaborations), hep-ex/0005030.

\bibitem{matematici}
M.~Daumens and P.~Minnaert, {\it J. Math. Phys.} {\bf 17} (1976) 2085.

\bibitem{Mathematica}
S.~Wolfram, {\em The Mathematica book},
3rd ed.\ (Wolfram Media/Cambridge Univ. Press, 1996).

\end{thebibliography}
\end{document}